\documentclass{article}
\usepackage{graphicx} 
\usepackage[utf8]{inputenc}
\usepackage{hyperref}
\usepackage{amsmath, amssymb}
\usepackage{caption}

\usepackage[backend=biber, style=chicago-authordate, natbib=true]{biblatex}
\title{High-Frequency Options Trading}
\author{Sid Bhatia}
\date{August 2024}

\usepackage[
    top=2cm,
    bottom=2cm,
    left=2cm,
    right=2cm,
    headheight=17pt, 
    includehead,
    includefoot,
    heightrounded, 
]{geometry}

\usepackage[backend=biber, style=chicago-authordate, natbib=true]{biblatex}
\begin{document}

\begin{titlepage}
    \centering
    Stevens Institute of Technology \\
    \vspace{1.5cm}
    \vspace{4cm}
    {\huge\bfseries High-Frequency Options Trading \\}
    \vspace{0.5cm}
    {\Large With Portfolio Optimization \\}
    \vspace{0.5cm}
    \vfill
    \textsc{\Large Sid Bhatia \\}
    \vfill
    {\large August 10th, 2024}
\end{titlepage}

\newgeometry{
    margin=0.8in
}

\newpage
\tableofcontents
\newpage

\section{Introduction}

\

In the rapidly evolving world of finance, algorithmic trading has emerged as a critical tool for investors seeking to maximize returns while managing risks efficiently \citep{hernandez2001optimal, hasbrouck1993highfrequency}. Within this domain, options trading, characterized by its complexity and potential for high returns, presents a unique set of challenges and opportunities \citep{black1973pricing, hull2018options}.

The core objective of our study is to investigate whether a high-frequency options trading strategy, augmented with advanced portfolio optimization techniques, can consistently generate positive returns \citep{cartea2015algorithmic}. This approach stands in contrast to traditional strategies that often rely on more straightforward methods such as simple long or short positions on options. Our strategy is intricate; it involves selecting a specific number of options from a larger pool based on predefined criteria such as the highest or lowest implied volatility \citep{gatheral2006volatility}. These selected options are then subjected to various trading strategies, ranging from basic long and short positions to more sophisticated methods involving portfolio optimization techniques like Standard Mean-Variance, Robust Methods, and more \citep{markowitz1952portfolio, fabozzi2007robust}.

The research is grounded in a comprehensive analysis of SPY options data sourced from Refinitiv. Our data set includes detailed information on calls and puts across different strike prices and expiry dates, recorded in five-minute intervals over a one-month period. This extensive data allows us to calculate critical metrics such as Option Greeks, implied volatility, and intrinsic and extrinsic values, providing a robust foundation for our analysis \citep{haug2007complete}. This research contributes to the expanding field of algorithmic trading by offering insights into the efficacy of advanced, high-frequency options trading strategies in dynamic and often unpredictable financial markets.

\section{Motivation}

\

The motivation behind our exploration into high-frequency options trading strategies, particularly those augmented by sophisticated portfolio optimization techniques, is rooted in both academic curiosity and the pragmatic nuances of financial markets. A primary driver is the discernible gap in existing literature regarding the application of high-frequency trading (HFT) strategies within options markets, especially when interwoven with advanced portfolio techniques. This research aims to bridge this gap, potentially offering new perspectives in financial engineering. Furthermore, the options markets, characterized by their complexity and rapid dynamics, often present unique inefficiencies. Our investigation is spurred by the possibility of identifying and exploiting these inefficiencies, leading to the development of innovative strategies that could provide a competitive edge in options trading.

The advancement in computational finance, bolstered by the surge in high-frequency financial data availability, has revolutionized financial analysis, presenting an opportune landscape for our research. By leveraging these technological advancements, we aim to explore new territories in option trading, utilizing sophisticated computational tools to dissect and understand complex market data. Moreover, the inherent nature of options as risk management tools aligns with our objective to not only seek high returns but also to contribute to more robust portfolio management practices, particularly in volatile market conditions. This aspect is especially pertinent given financial events such as the 2010 Flash Crash, which underscore the necessity for strategies that maintain efficacy even under extreme market volatility. 

\section{Methodology}

\

Using our data, we first split the data into two subsets: one with little to no null values to represent more frequently traded options, and the other with anywhere from one to three null values, indicating less frequently traded options. We added certain features necessary for our analysis, such as each contract’s Spot Price ($S_0$), Strike ($K$), and Time to Maturity (in years, $T$), so that we can calculate each option’s Implied Volatility and Greeks.

Because we are analyzing American options, it may not be accurate to use the Black-Scholes formula, as it's a closed-form solution for the price of European Options. Therefore, for our analysis, we use the Binomial Tree model with a sufficiently large number of time-steps to improve the accuracy of the price estimate, and we use the formulas for the Greeks of American Options as presented by Muroi and Suda (2017).

From there, we constructed different strategies based on certain criteria. For example, in each subset, we would create an investment universe of options by selecting the three options with the highest implied volatility (IV) and the three with the lowest, repeating this process for each of the option Greeks as well as combinations of Greeks. We then compared the performances of different portfolio optimization models for each dataset: a dynamic algorithm in which new weights are generated every certain number of time-steps and four static algorithms, namely the standard Markowitz model, a portfolio with a risk-free rate, a portfolio with shrinkage, and a robust optimization method.

\section{Data}

\

We utilized comprehensive datasets sourced from Refinitiv, accessed via the Hanlon Financial Systems Lab. Our primary focus was on SPY options, which included both calls and puts. The data was recorded at 1-hour intervals, covering a broad range of strike prices, and spanned from October 1, 2023, to November 1, 2023. The dataset for this period is extensive, comprising 21 columns and 6,214,874 rows. Additionally, we analyzed a dataset encapsulating the events of May 6, 2010, known as the Flash Crash. This dataset featured SPY options at 5-minute intervals, also across varying strike prices. The Flash Crash data structure consisted of 21 columns and 552,672 rows.

In preparing the data for analysis, we carefully addressed columns with a significant number of missing values (NA's), identified and rectified outliers, and standardized the data to ensure consistency across the datasets. The cleaning process resulted in a refined dataset with the following columns: '\#RIC', 'Domain', 'Date-Time', 'Type', 'Open', 'High', 'Low', 'Last', 'Volume', 'No. Trades', 'Open Bid', 'High Bid', 'Low Bid', 'Close Bid', ‘No. Bids', 'Open Ask', 'High Ask', 'Low Ask', 'Close Ask', 'No. Asks', 'Mid Open', 'Mid Close', 'Root', 'Strike Price', 'Maturity', and 'Contract Type'.

To facilitate a nuanced analysis, we segregated the options into two categories based on liquidity: liquid options, characterized by having 0 to 1 missing values (NA's), and illiquid options, with 3 to 7 missing values (NA's). This distinction allowed us to examine the performance of trading strategies under varying liquidity conditions, and we maintained separate CSV files for each category accordingly.

\section{Implied Volatility}

\

To calculate the implied volatility, (IV) we utilize the Newton-Raphson algorithm, which is designed to find the roots of a real-valued function. We first express the option price as a function of volatility so that we are able to take its derivative. Then we start with an initial guess of what the volatility could be and iteratively update this guess as shown below:

\begin{equation}
x_{n+1} = x_{n} - \frac{f(x_{n})}{f'(x_{n})}
\end{equation}

We repeat this process until the market price matches the Binomial Tree price, which is used as a numerical approximation with a large number of time steps. 

After we have found the IV for each contract, we can construct some investment universe based on this by taking the three options with the highest IV and the three with the lowest, for example. By sub-setting our options data in this way, we can create some basic initial strategy; for example, longing the top 3 and shorting the bottom 3, or a more robust approach in which we calculate portfolio weights periodically. 

The basic constraints set for the problem are that all weights must add to 1 and that each must be between 1\% and 40\% of the total portfolio. An additional constraint that we add later on is a maximum IV, meaning that the portfolio IV, defined as the weighted sum of individual IVs, cannot exceed this amount. Some results are shown below:

\section{Option Greeks}

\subsection{Basic}
To calculate the Greeks, we utilized $\Delta MSA(s,0)$ for American Options. This is specific to American options, which have the feature of early exercise. It considers the possibility of early exercise and adjusts the delta calculation accordingly. It involves the partial derivative of the payoff function with respect to the stock price, taking into account the expected future option value under random market movements.

\begin{equation}
    \Delta_A^{M S}(s, 0)= \begin{cases}
    \partial_s \Phi(s) & \text{if } s \in \mathcal{S}_0 \\
    \frac{e^{-r \Delta t}}{s \sigma \Delta t} \mathbb{E}\left[\mathcal{A}\left(s e^{\sigma \epsilon_1}, \Delta t\right)\left(\epsilon_1-\mu \Delta t\right)\right] & \text{if } s \in \mathcal{C}_0
    \end{cases}
\end{equation}

While there are similarities in the use of derivatives and the influence of expected future values, $\Delta MSA(s,0)$ is designed to handle the additional complexity introduced by the early exercise feature of American options, making it distinct from the classic Black-Scholes delta.

Gamma ($\Gamma$) for American options represents the second partial derivative of the option price with respect to the underlying asset's price. In contrast to European options, American options allow for early exercise, introducing complexity into the gamma calculation. To adapt, a numerical approach like the binomial tree or finite difference methods is employed.

Theta ($\Theta$) for American options is determined by calculating the time decay of the option's value over a small time interval. Similar to gamma, a numerical approach is often employed due to the potential for early exercise. The calculation involves perturbing the time to expiration, recalculating option values, and then using finite differences to estimate the first derivative with respect to time. This process allows for an approximation of how much the option's value changes as time progresses, providing insights into the impact of time decay on the option premium.

After conducting these calculations, we tested the returns of strategies going long the three options with the highest delta, gamma, or theta, and shorting the three with the lowest. We also created a dynamic portfolio with the basic constraints set for the problem that all weights must add to 1 and that each must be between 1\% and 40\% of the total portfolio as well as be rebalanced every period. The following show the results of these two strategies:

\begin{figure}[ht]
  \centering
  \begin{minipage}[b]{0.45\textwidth}
    \includegraphics[width=\textwidth]{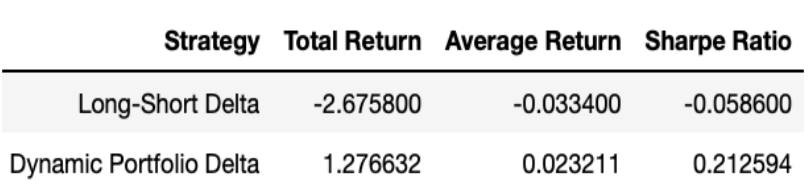}
    \captionsetup{justification=centering}
    \caption{Delta Portfolio Analysis}
    \label{fig:delta_port}
  \end{minipage}
  \hfill 
  \begin{minipage}[b]{0.45\textwidth}
    \includegraphics[width=\textwidth]{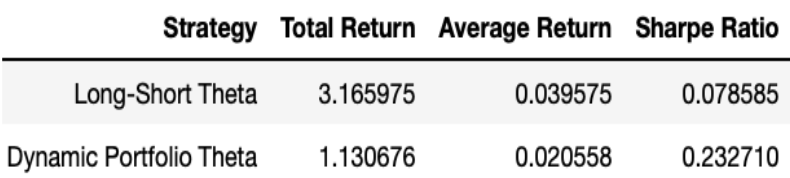}
    \captionsetup{justification=centering}
    \caption{Theta Portfolio Analysis}
    \label{fig:theta_port}
  \end{minipage}
\end{figure}

\subsection{Advanced} 

Building on the foundational understanding of the basic Greeks and their role in American option pricing, our analysis progressed to the application of advanced Greeks in dynamic portfolio strategies. The advanced Greeks—Vega ($\nu$) and Rho ($\rho$) extend the dimensionality of risk assessment by considering not just the immediate impact of price and time but also the influence of volatility and interest rates on the option's price.

The dynamic portfolio strategies, incorporating Delta \& Rho, Delta \& Vega, Vega \& Rho, and Delta \& Gamma interactions, provided an enriched perspective on risk management. These advanced Greeks interactions were analyzed through a range of market conditions, reflecting the multifaceted nature of option valuation in a real-world setting.

Delta \& Rho and Delta \& Vega strategies were instrumental in balancing the immediate price movement risks with the more subtle influences of volatility and interest rates. The Vega \& Rho combination offered a unique perspective on how volatility and economic policy shifts can impact option pricing. The Delta \& Gamma strategy emphasized the importance of managing the convexity in price movements, particularly beneficial in markets exhibiting large directional trends.

\begin{figure}[ht]
  \centering
  \begin{minipage}[b]{0.45\textwidth}
    \includegraphics[width=\textwidth]{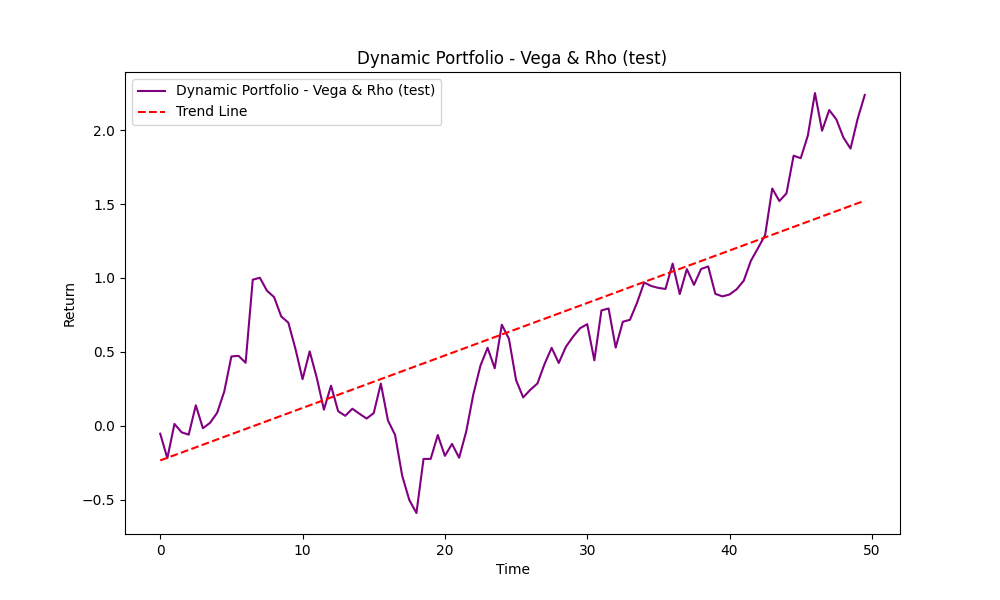}
    \captionsetup{justification=centering}
    \caption{Dynamic Portfolio - Vega \& Rho Strategy}
    \label{fig:delta_rho}
  \end{minipage}
  \hfill 
  \begin{minipage}[b]{0.45\textwidth}
    \includegraphics[width=\textwidth]{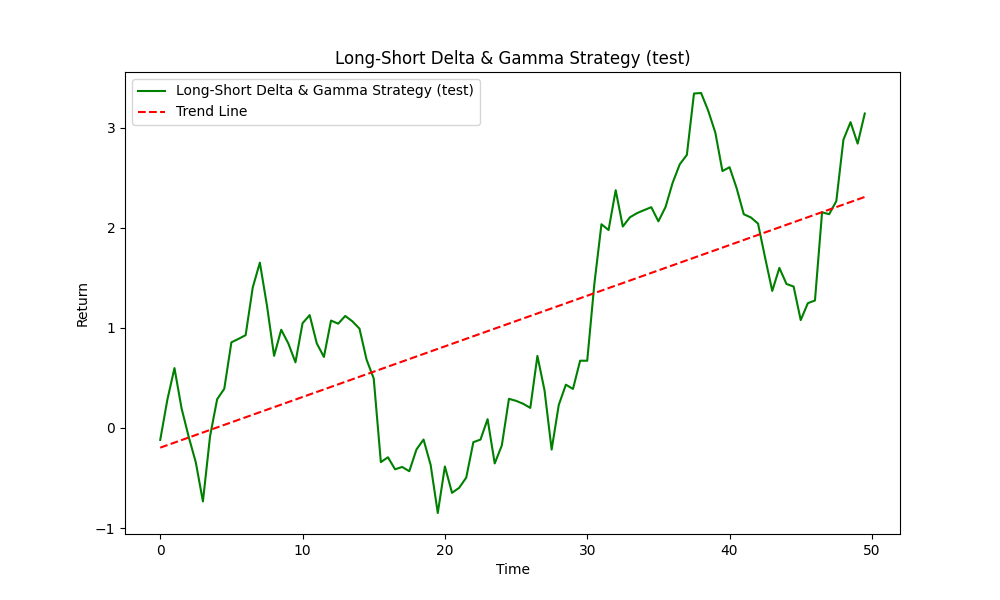}
    \captionsetup{justification=centering}
    \caption{Long-Short Delta \& Gamma}
    \label{fig:delta_vega}
  \end{minipage}
\end{figure}

\begin{figure}[ht]
  \centering
  \begin{minipage}[b]{0.45\textwidth}
    \includegraphics[width=\textwidth]{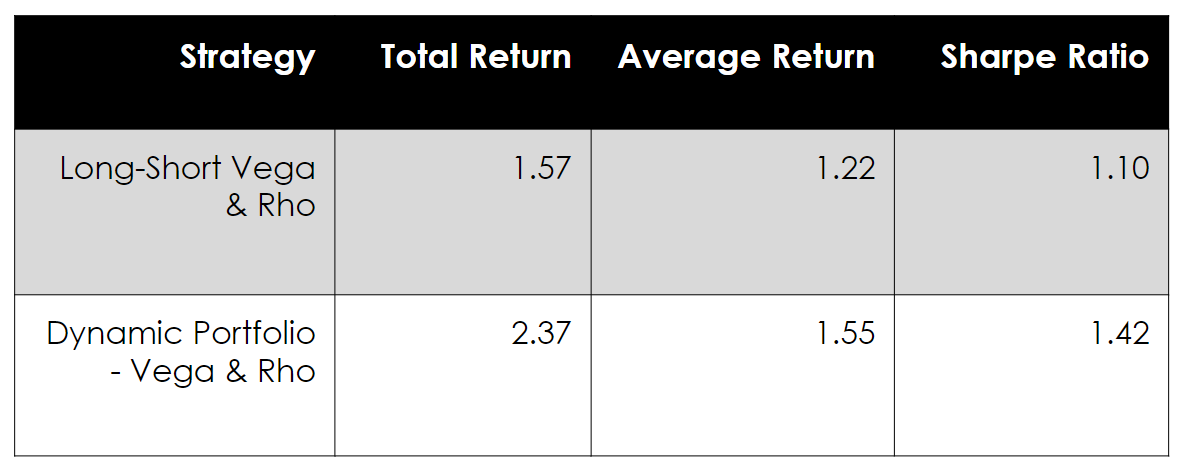}
    \captionsetup{justification=centering}
    \caption{Vega \& Rho Portfolio Analysis}
    \label{fig:delta_rho}
  \end{minipage}
  \hfill 
  \begin{minipage}[b]{0.45\textwidth}
    \includegraphics[width=\textwidth]{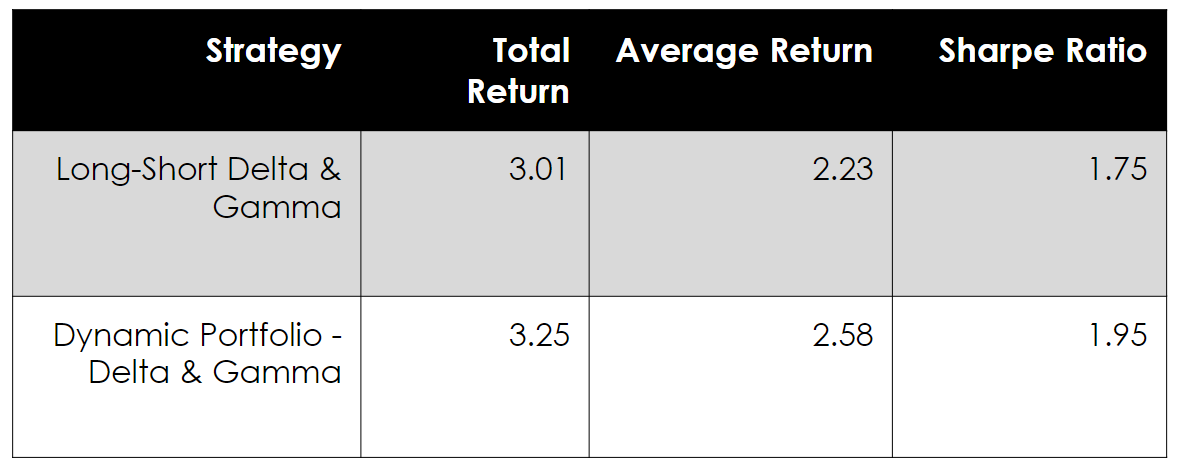}
    \captionsetup{justification=centering}
    \caption{Delta \& Gamma Portfolio Analysis}
    \label{fig:delta_vega}
  \end{minipage}
\end{figure}

Our analysis demonstrated that each strategy's effectiveness varies with market dynamics. For instance, the Delta \& Rho strategy showed vulnerability to interest rate changes, while the Delta \& Vega strategy highlighted opportunities within volatile market environments. The Vega \& Rho strategy navigated the dual forces of volatility and interest rate shifts, whereas the Delta \& Gamma strategy sought to capitalize on the momentum of price movements.

\section{Standard Portfolio Strategies}

To have a reference to compare the performance of our basic long-short and dynamic portfolio strategies, we tested four different static portfolio strategies, in which only one set of weights is generated for the entire time horizon with a certain target return in mind. Unlike for the dynamic optimization with the implied volatility or the Greeks, there were no additional constraints set upon the portfolio weights other than that they had to sum to one. 

\section{Conclusion/Results}
\

In conclusion, our investigation into high-frequency options trading strategies augmented with sophisticated portfolio optimization techniques has yielded insightful findings. Our analysis suggests that while basic long-short strategies involving implied volatility and Greeks generally tend to underperform, certain strategies centered around Theta, Rho, and combined Greeks show promise. These strategies, particularly when they incorporate advanced portfolio optimization techniques, demonstrate a capacity to navigate the complexities and rapid dynamics of the options market effectively.

Further, the exploration of different portfolio strategies, ranging from static models to dynamic algorithms, has revealed nuanced understandings of market behavior and strategy efficacy. The dynamic nature of the high-frequency trading environment underscores the importance of adaptable and responsive strategies.

Future research could benefit from a deeper exploration into the optimization of strategy parameters (hyperparameter optimization), such as adjustments in the number of options in the long-short strategy, implementation of minimum and maximum weight constraints, and variations in rebalancing frequency. Additionally, examining less frequently traded options or incorporating historical data could provide a more comprehensive view of the strategies’ performance in different market conditions. These insights could be pivotal in refining existing strategies or developing new approaches.

\newpage

\section{References/Tools}

\

\printbibliography

\end{document}